\documentclass[preprint,12pt]{elsarticle}
\usepackage{paper-shay-IPL}

\usepackage{amssymb}

\begin{document}

\begin{frontmatter}
\title{Smoothness of Schatten Norms and Sliding-Window \\ Matrix Streams}
% \title{Sliding-Window Streams: Schatten Norms are as Smooth as $L_p$-Norms}

\author{Robert Krauthgamer}\ead{robert.krauthgamer@weizmann.ac.il}
\author{Shay Sapir}\ead{shay.sapir@weizmann.ac.il}

\address{Weizmann Institute of Science}

\begin{abstract}

    Large matrices are often accessed as a row-order stream.
    We consider the setting where rows are time-sensitive (i.e. they expire),
    which can be described by the sliding-window row-order model,
    and provide the first $(1+\epsilon)$-approximation of Schatten $p$-norms in this setting.
    Our main technical contribution is a proof that Schatten $p$-norms in row-order streams are smooth, and thus fit the smooth-histograms technique of Braverman and Ostrovsky (FOCS 2007) for sliding-window streams.

\end{abstract}

\begin{keyword}
%% keywords here, in the form: keyword \sep keyword
Sliding window streams \sep Matrix streams \sep Schatten norms \sep Smooth histograms
\end{keyword}
\end{frontmatter}

\section{Introduction}
Many modern data-sets are viewed as matrices that have millions or even billions of dimensions. 
Access to such large matrices is often done in a streaming fashion,
meaning that the input is a sequence of items that can be read only sequentially, usually in one pass.
Algorithms in this model are restricted to have small memory compared to the input size, and cannot access earlier input portions, hence they effectively have to compress the input. 

In matrix streams, three settings are often considered.
In all of them, there is an underlying matrix $A\in\R^{N\times m}$ initialized to the all-zeros matrix.
In the row-order model, the input is a stream of non-zero entries of $A$ presented in row-order, hence we regard it as if the input is a stream of rows of $A$, i.e. vectors in $\mathbb{R}^m$ (see e.g. \citet{liberty2013simple}).%
\footnote{The definition as a stream of entries (and not rows) avoids working space linear in the size of the row, which is necessary to process an entire row.}
In the entry-wise model, the input is a stream of non-zero entries of $A$ in arbitrary order, and
in the turnstile model, the input is a stream of additive updates to the entries of $A$, where these updates can also be negative.

Scenarios where the input is time-sensitive are often modeled by a sliding-window, meaning that at any point in time, the input is only the most recent $n$ stream items (earlier items are ignored) called the active window \citep{datar2002maintaining}.
Particularly, in row-order matrix streams, the input is a matrix made of the most recent $n$ rows (and not entries).
Several linear-algebra problems were addressed 
in this model,
e.g. covariance matrix approximation, PCA \citep{wei2016matrix,braverman2020near}, spectral approximation, $\ell_1$ subspace embedding and row/column subset selection \citep{braverman2020near}.

The spectrum of a matrix contains information related to many properties of the matrix (e.g, rank, condition number), and computing it is often a first step in data analysis. 
We focus on fundamental functions of the spectrum, called Schatten norms.
Formally, for $p\geq 1$, the \emph{Schatten $p$-norm} of a matrix $A\in\R^{n\times m}$ is
$$
\|A\|_{S_p} = (\sum_i \sigma_i^p)^{1/p},
$$
where $\sigma_1\geq\sigma_2\geq...\geq \sigma_{\min \{n,m\}}\geq 0$ are the singular values of $A$.
This definition extends to $0\leq p<1$, for which it is not a norm.
Special cases include $p=0,1,2,\infty$, which are the rank, the trace norm, the Frobenius norm and the spectral/operator norm, respectively.
Schatten norms can be used to estimate spectral-sum functions using Chebyshev polynomials, as explained in \citep{khetan2019spectrum}; 
or to estimate the spectrum itself via the method of moments, as explained in \citep{kong2017spectrum};
or for matrix completion \citep{nie2012low}. 
There is a long line of work on computing Schatten $p$-norms in a stream, see \citep{li2014sketching,li2016approximating,li2016tight, braverman2018matrix,braverman2020schatten}, and the further  motivation provided therein.

We focus on approximating the Schatten $p$-norm of a sliding-window row-order stream.
Previous work on this problem achieved $(\sqrt{2}+\epsilon)$-approximation \citep{krauthgamer2019almost}, and as explained next, we provide the first $(1+\epsilon)$-approximation.\footnote{A $c$-approximation to $x\in\R$ is $y\in\R$ such that $x\leq y\leq c x$ (if $y$ is a random variable, then it should hold with high probability).}

\subsection{Technical Contribution}

A key algorithmic approach for sliding-window streams is the smooth-histograms technique of \citet{braverman2007smooth}.
It is applicable to functions that (i) satisfy a certain smoothness criterion,
and (ii) admit a one-pass $(1+\epsilon)$-approximation algorithm.
The idea is to dynamically
maintain several instances of that algorithm on different suffixes of the stream,
altogether achieving $(1+\epsilon)$-approximation on the sliding-window \citep{braverman2007smooth}.
Our main technical contribution is simply to prove that Schatten $p$-norms satisfy the smoothness criterion.
An immediate corollary using the smooth-histograms technique of \citet{braverman2007smooth} is  $(1+\epsilon)$-approximation algorithm for Schatten $p$-norm in the sliding-window row-order model, whose space requirement is comparable to the row-order one-pass (not sliding-window) setting.
In contrast, previous work by
\citet{krauthgamer2019almost}  relied on
relaxing the smoothness criterion of \citep{braverman2007smooth}, 
which captures a more general family of functions, at the cost of a
worse approximation factor $O(1)$.

% We achieve $(1+ \epsilon)$-approximation for Schatten $p$-norms in row-order sliding-window streams, 
% using space that is comparable to the row-order one-pass (not sliding-window) setting.
% We use the smooth-histogram technique of \citet{braverman2007smooth} and our main technical contribution is to prove that the Schatten $p$-norms are smooth in the row-order setting.

Let us recall the smooth-histograms technique, instantiating it for our context of row-order matrix streams.
Let $X,Y,A$ and $C$ denote real matrices with $m$ columns and at most $n$ rows ($n$ is the size of the window), and interpret them also as row-order streams.
Using this notation, the smoothness definition of \citet[Definition 1]{braverman2007smooth} reads as follows.

\begin{definition}\label{def:smoothness}
A real-valued function $f$ defined on real matrices is called $(\alpha,\beta)$-smooth, where $0<\beta \leq \alpha<1$, if it satisfies the following.
\begin{enumerate}
    \item Non-negative: for every real matrix $A$, it holds that $f(A)\geq 0$.
    \item Non-decreasing: 
    for all $A=\big[\begin{smallmatrix}X \\ Y\end{smallmatrix}\big]$, it holds that $f(A)\geq f(X)$.
    \item Bounded: for every matrix $A$ with integral entries bounded by $\poly(n)$, it holds that $f(A)\leq \poly(n)$.
    \item Smooth: for all real matrices $A=\big[\begin{smallmatrix}X \\ Y\end{smallmatrix}\big]$ and $C$, if $(1-\beta)f(A)\leq f(Y)$ then $(1-\alpha)f\big(\big[\begin{smallmatrix}A \\ C\end{smallmatrix}\big]\big)\leq f\big(\big[\begin{smallmatrix}Y \\ C\end{smallmatrix}\big]\big)$.
\end{enumerate}
\end{definition}

Assume that the entries of the matrices are integers bounded by $\poly(n)$.

\begin{theorem}\label{lem:smooth_thm}\citep[Theorem 3]{braverman2007smooth}
Let $f$ be an $(\alpha,\beta)$-smooth function. 
If there is a one-pass algorithm $\Lambda$ that maintains $(\epsilon,\delta)$-approximation of $f$ on a stream
using $g(\epsilon,\delta)$ bits of space and performing $h(\epsilon,\delta)$ operations per stream item, 
then there exists an algorithm $\Lambda'$ that maintains $(\alpha+\epsilon,\delta)$-approximation of $f$ in sliding-window streams
using $O(\tfrac{1}{\beta}(g(\epsilon,\tfrac{\delta\beta}{n\log n}) + \log n) \log n)$ bits of space and $O(\tfrac{1}{\beta}h(\epsilon,\tfrac{\delta\beta}{n\log n})\log n)$ operations per item.%
\footnote{An $(\epsilon,\delta)$-approximation of $f$ refers to
  a random variable that with probability at least $1-\delta$
  is a $(1+\epsilon)$-approximation of $f$.}%
\textsuperscript{,}%
\footnote{Theorem 3 in \cite{braverman2007smooth} has a minor issue in the parameter settings. Although there are $O(\tfrac{1}{\beta}\log n)$ algorithms in the histogram at any fixed time, we might need correctness of all algorithms at every time-step to avoid adversarial failures. Hence the probability of failure needs to be $\tfrac{\delta\beta}{n\log n}$ rather than $\tfrac{\delta\beta}{\log n}$.}
\end{theorem}

We can now state our main technical result, that Schatten $p$-norms are smooth. Our proof generalizes the proof of \citet[Lemma 4]{braverman2007smooth} that $\ell_p$-norms are smooth, using matrix analysis tools that are based on pinching and monotonicity of the trace function.

\begin{restatable}{proposition}{largep}\label{lem:p>2}
For all $p\geq 2$ and $0<\epsilon<1$, the Schatten $p$-norm 
is $(\epsilon,\frac{\epsilon^{p/2}}{p/2})$-smooth.
\end{restatable}
\begin{restatable}{proposition}{smallp}\label{lem:p<2}
For all $0<p\leq 2$ and $0<\epsilon<1$, the Schatten $p$-norm 
is $(\epsilon,\epsilon)$-smooth.
\end{restatable}

These two propositions provide the same smoothness bound for $p=2$, which is just the Frobenius norm. In the row-order model, computing the Frobenius norm is equivalent to computing the sum of positive numbers, which in turn is known to be $(\epsilon,\epsilon)$-smooth \citep{braverman2007smooth}.

Our smoothness bounds match the known bounds for $\ell_{p/2}$-norms of vectors under insertions (no deletions) \citep[Lemma 4]{braverman2007smooth}, i.e. where the input is a stream of items $(i,\delta)\in [m]\times \R_+$, representing updates of the form $x_i\leftarrow x_i + \delta$ to a vector $x\in\R_+^m$.
Further, our bounds are more general, since
$\ell_{p/2}$-norms in this model can be simulated by Schatten $p$-norms in row-order streams, as follows. 
For an initial vector $x\in\R_+^m$, let $A\in\R^{m\times m}$ be a diagonal matrix with entries $\{\sqrt{x_i}\}_i$ on its diagonal.
Then, $A^\top A$ is a diagonal matrix having the entries of $x$ on its diagonal and $\|A\|_{S_p}^2 =\|A^\top A\|_{S_{p/2}}=\|x\|_{p/2}$.
Maintain this property of $A^\top A$, as follows.
For an item $(i,\delta)$ with the update $x_i\leftarrow x_i +\delta$, where $\delta>0$,
update $A \leftarrow \big[\begin{smallmatrix} A \\ \sqrt{\delta}e_i^\top \end{smallmatrix}\big]$, where $e_i$ is the $i$-th standard basis vector.
Hence, the update to $A^\top A$ is $A^\top A \leftarrow A^\top A + \delta e_i e_i^\top$. 
Thus, the update to the diagonal of $A^\top A$ is the same as the update to $x$. 
This reduction shows that our smoothness bounds generalize the bounds of \citet[Lemma 4]{braverman2007smooth} for $\ell_{p/2}$-norms
(they proved that $\ell_q$-norm is $(\epsilon,\epsilon^q/q)$-smooth for $q\geq 1$, and $(\epsilon,\epsilon)$-smooth for $0<q<1$).

\subsection{Main Results}

Our main result follows immediately from Propositions \ref{lem:p>2} and \ref{lem:p<2} using the smooth-histograms technique (Theorem \ref{lem:smooth_thm}). 
It shows that any one-pass algorithm to $(1+\epsilon)$-approximate the Schatten $p$-norm in a row-order stream, implies a sliding-window algorithm with almost the same space complexity.

\begin{corollary}\label{cor:main_reduction}
Let $\Lambda$ be a one-pass algorithm that maintains $(\epsilon,\delta)$-approximation of Schatten $p$-norm in row-order stream using $g(\epsilon,\delta)$ bits of space and performing $h(\epsilon,\delta)$ operations per stream item (i.e. a matrix row). 
Then there exists an algorithm $\Lambda'$ that maintains  $(2\epsilon,\delta)$-approximation of Schatten $p$-norm in sliding-window row-order streams such that:

\begin{enumerate}
    \item if $p\geq 2$, it uses $O(\frac{p}{\epsilon^{p/2}}(g(\epsilon,\frac{2\delta\epsilon^{p/2}}{p n\log n}) + \log n) \log n)$ bits of space
        and 
    $O(\frac{p}{\epsilon^{p/2}}h(\epsilon, \frac{2\delta\epsilon^{p/2}}{p n\log n}) \log n)$ 
    operations per stream item.
    \item if $p<2$, it uses $O(\frac{1}{\epsilon}(g(\epsilon,\frac{\delta\epsilon}{n\log n}) + \log n) \log n)$ bits of space
        and 
    $O(\frac{1}{\epsilon}h(\epsilon,\frac{\delta\epsilon}{n\log n})\log n)$ 
    operations per stream item.
\end{enumerate}
\end{corollary}

Thus, every one-pass $(1+\epsilon)$-approximation algorithm for Schatten $p$-norm in row-order stream implies a sliding-window $(1+\epsilon)$-approximation algorithm with similar space requirement. 
It remains open whether the overhead can be avoided.
For $\ell_p$-norms, this overhead was recently removed by \citet{DBLP:journals/arxiv/WoodruffZ20}, who developed a new framework for sliding-window streams, which applies to smooth functions that admit a certain type of algorithms, called difference estimator.
If one were to design difference estimators for Schatten $p$-norms, then 
% by Propositions \ref{lem:p>2} and \ref{lem:p<2} and Framework 1.7 of \citep{DBLP:journals/arxiv/WoodruffZ20}, 
this will imply almost no overhead.

Unfortunately, for $p<2$ no non-trivial algorithm is known for row-order streams; in fact,
every one-pass $(1+\epsilon)$-approximation of Schatten $p$-norm of $n\times n$ matrices in row-order streams must use at least $\Omega(n^{1-g(\epsilon)})$ bits of space, where $g(\epsilon)\to 0$ as $\epsilon\to 0$ \citep[Theorem 5.3 in arXiv version]{braverman2018matrix}, and
this lower bound extends immediately to the more restricted sliding-window model.

For $p>2$, there are two known one-pass $(\epsilon,\delta)$-approximation algorithms for Schatten $p$-norm of $n\times n$ matrices in row-order streams:
% (i) for \emph{Schatten $4$-norm}, using $O(\epsilon^{-2} \log n)$ bits of space \citep[Theorem 7.2]{braverman2020schatten};
(i) for $O(1)$-\emph{sparse matrices and even integer $p\geq 4$},
using $\tilde{O}_p(n^{1-4/\ceil{p}_4} \poly(\epsilon^{-1}))$ bits of space \citep[Theorem 6.1 and Section 1.3 in arXiv version]{braverman2018matrix} (improving over \cite{li2016approximating});%
\footnote{A matrix is said to be sparse if there are $O(1)$ non-zero entries in every row/column. We use $\ceil{p}_4$ to denote the smallest multiple of $4$ that is larger or equal to $p$, and similarly $\floor{p}_4$ to denote the largest multiple of $4$ that is smaller or equal to $p$.
}
and (ii) for \emph{even integer $p\geq 4$} using $\tilde{O}_p(\epsilon^{-2} n^{2-\frac{8}{p}})$ bits of space \cite[Theorems 3.3 and 3.8]{braverman2018matrix} (improving over \cite{li2014sketching}).%
\footnote{The algorithms in \cite[Theorems 3.3 and 3.8]{braverman2018matrix} are for integer $p\geq 2$ and PSD matrices in turnstile streams, and require space $\tilde{O}_{p}(\epsilon^{-2}n^{2-4/p})$. 
  As mentioned in \cite{braverman2018matrix}, given an even integer $p$ and a matrix $A$ in row-order,
  one can apply these algorithms to the PSD matrix $B=A^\top A$ and $p'=p/2$,
  to estimate $\|B\|_{S_{p'}} = \|A^\top A\|_{S_{p/2}} = \|A\|_{S_{p}}^2$.
  The updates do not require additional space, since these algorithms only rely on bilinear sketches (i.e., sketches of the form $G_1^\top B G_2 = G_1^\top A^\top A G_2$).
}
For $p=4$, \citet{braverman2020schatten} provide a simple and explicit algorithm with the same space bound.

These bounds are summarized in Table \ref{table:SchattenSpaceBounds}, together with the corresponding overhead of Corollary \ref{cor:main_reduction} for sliding-window streams.
These algorithms are stated for $n\times n$ matrices, but they immediately generalize to $n\times m$ matrices with $n>m$, and the polynomial dependence in the space bound is with respect to $m$ (and not $n$). 
Plugging these algorithms into Corollary \ref{cor:main_reduction}, we get the following.

\begin{corollary}
There are algorithms that maintain $(\epsilon,\delta)$-approximation of Schatten $p$-norm in sliding-window row-order streams with the following space requirements.
\begin{enumerate}
    % \item for $p=4$: using $O(\epsilon^{-4}\log^2 n \log \tfrac{n\log n}{\delta \epsilon})$ bits of space.
    \item for even integer $p\geq 4$ and $O(1)$-sparse matrices: using $\tilde{O}_p(n^{1-4/\ceil{p}_4}\poly(\epsilon^{-1}))$ bits of space.
    \item for even integer $p\geq 4$: using $\tilde{O}_p(\epsilon^{-2-\frac{p}{2}} n^{2-\frac{8}{p}})$ bits of space.
\end{enumerate}
\end{corollary}

Previously, only $(\sqrt{2}+\epsilon)$-approximation was known (with similar space requirement) \cite{krauthgamer2019almost}.

\renewcommand{\arraystretch}{1.25}
\begin{table*}[!t]
\caption{\label{table:SchattenSpaceBounds} Known space bounds (in bits) for one-pass $(1+\epsilon)$-approximation of Schatten $p$-norms in the row-order model and the corresponding multiplicative overhead in the sliding-window model.
} 
\begin{center}
\begin{tabulary}{\textwidth}{|c  |c c| c|}
\hline
Which $p>0$                         & One-Pass and Row-Order &                                             &  Sliding-Windows Overhead\\ 
\hline
\hline
% $p=4$                           & $\tilde{O}(\epsilon^{-2})$  & \citep{braverman2020schatten}  & $\tilde{O}(\epsilon^{-2})$ \\ 
even $p\geq 4$, sparse matrix   & $\tilde{O}_{p,\epsilon}(n^{1-\frac{4}{\ceil{p}_4}} )$ &\citep{li2016approximating,braverman2018matrix} & $\tilde{O}_p(\epsilon^{-\frac{p}{2}})$ \\ 
even $p\geq 4$, every matrix    & $\tilde{O}_{p,\epsilon}(n^{2-\frac{8}{p}})$ & \citep{li2014sketching,braverman2018matrix} & $\tilde{O}_p(\epsilon^{-\frac{p}{2}})$ \\ 
\hline
$p$ not even                    & $\Omega(n^{1-g(\epsilon)})$ & \citep{braverman2018matrix} & - \\
even $p\geq 4$, sparse matrix   & $\Omega(n^{1-\frac{4}{\floor{p}_4}})$ & \citep{braverman2020schatten} & - \\
\hline
\end{tabulary}
\end{center}
\end{table*}

%%% Local Variables:
%%% mode: latex
%%% TeX-master: "main"
%%% End:

\section{Smoothness of Schatten Norms for $p\geq 2$}
In this section, we prove Proposition \ref{lem:p>2}, 
that Schatten $p$-norms for $p\geq 2$ in row-order streams are $(\epsilon,\tfrac{\epsilon^{p/2}}{p/2})$-smooth.
\citet{krauthgamer2019almost} showed that Schatten norms are non-negative, non-decreasing and bounded. 
We complete this observation, and analyze the non-trivial property of Definition \ref{def:smoothness}. 
Our proof is based on the proof for $\ell_q$ frequency moments for $q>1$ \citep[Lemma 4]{braverman2007smooth}.
We will need two auxiliary lemmas, as follows.

The first lemma has a simple proof using a pinching technique.
It is given as an exercise by \citet{bhatia1997matrix}, and we provide its proof for completeness.

\begin{lemma}\label{fact:SchattenIneqForSum}\citep[Problem II.5.4.]{bhatia1997matrix}
For all $p\geq 2$, real matrices $X,Y$ with the same row length and $A=\big[\begin{smallmatrix}X\\Y\end{smallmatrix}\big]$,
\[
\|A\|_{S_p}^p = \|X^TX+Y^TY\|_{S_{p/2}}^{p/2}\geq \|X^TX\|_{S_{p/2}}^{p/2} + \|Y^TY\|_{S_{p/2}}^{p/2}.
\] 
\end{lemma}

\begin{proof}
Note that
\[
\|A\|_{S_p}^p=\|A^TA\|_{S_{p/2}}^{p/2}=\|AA^T\|_{S_{p/2}}^{p/2}=
\bigg\|\bigg[\begin{matrix}
XX^T & XY^T\\
YX^T & YY^T
\end{matrix}\bigg]
\bigg\|_{S_{p/2}}^{p/2}.
\]
Now, denote $Z=\bigg[\begin{matrix}
XX^T & XY^T\\
YX^T & YY^T
\end{matrix}\bigg]$ and $U=\bigg[\begin{matrix}
I & 0\\
0 & -I
\end{matrix}\bigg]$. Then, $\|A\|_{S_p}^p = \|Z\|_{S_{p/2}}^{p/2}$ and 
$$\bigg[\begin{matrix}
XX^T & 0\\
0 & YY^T
\end{matrix}\bigg]=\frac{1}{2}(Z+UZU^T).$$
Thus, by the triangle inequality,
\begin{align*}
\|XX^T\|_{S_{p/2}}^{p/2} + \|YY^T\|_{S_{p/2}}^{p/2} &= 
\bigg\|\bigg[\begin{matrix}
XX^T & 0\\
0 & YY^T
\end{matrix}\bigg]\bigg\|_{S_{p/2}}^{p/2} \\
&\leq
\big(\tfrac{1}{2}(\|Z\|_{S_{p/2}}+\|UZU^T\|_{S_{p/2}})\big)^{p/2}
= \|Z\|_{S_{p/2}}^{p/2},
\end{align*}
where the last step holds since Schatten norms are unitarily invariant.
\end{proof}

The second lemma is a technical bound. % for the error.
\begin{lemma}\label{fact:epsBoundForLargeP}
For all $0<\epsilon\leq 1$ and $p>2$, it holds that $\sqrt{1-(2\epsilon^{p/2} - \epsilon^p)^{2/p}}\geq 1-\epsilon$.
\end{lemma}
\begin{proof}
We begin by analyzing the function $f(\epsilon)=(2-\epsilon)^{p/2}+\epsilon^{p/2}$. Its derivative is $$f'(\epsilon)=\tfrac{p}{2}(-(2-\epsilon)^{p/2-1}+\epsilon^{p/2-1})<0,$$ for the given range of $\epsilon$. Hence it is decreasing, and its minimum is at $\epsilon=1$, i.e. $f(\epsilon)\geq f(1)=2$. Hence,
\begin{align*}
    &(2-\epsilon)^{p/2}\geq 2-\epsilon^{p/2} \Longrightarrow   (2\epsilon-\epsilon^2)^{p/2}\geq 2\epsilon^{p/2}-\epsilon^{p} \\
    \Longrightarrow   &2\epsilon-\epsilon^2\geq (2\epsilon^{p/2}-\epsilon^{p})^{2/p} \Longrightarrow   \sqrt{1-(2\epsilon^{p/2}-\epsilon^{p})^{2/p}} \geq 1- \epsilon.
\end{align*}
\end{proof}

We are ready to prove Proposition \ref{lem:p>2}.

\begin{proof}[Proof of Proposition \ref{lem:p>2}]
Schatten norms are non-negative, non-decreasing and bounded \citep[Corollary 3.9]{krauthgamer2019almost}. 
Let $p\geq 2$ and $X,Y,C$ real matrices with rows of length $m$, as in Definition \ref{def:smoothness}, such that $A=\big[\begin{smallmatrix}X \\ Y\end{smallmatrix}\big]$ satisfies $(1-\frac{\epsilon^{p/2}}{p/2})\|A\|_{S_p}\leq \|Y\|_{S_p}$. 
Since $\|A\|_{S_p}^p = \|A^\top A\|_{S_{p/2}}^{p/2}$,
our goal is to prove that $(1-\epsilon)\|A^\top A + C^\top C\|_{S_{p/2}}^{1/2}\leq \|Y^\top Y + C^\top C\|_{S_{p/2}}^{1/2}$.
We have
\[
\|Y\|_{S_p}^p\geq 
\big(1-\tfrac{\epsilon^{p/2}}{p/2}\big)^p\|A\|_{S_p}^p
\geq (1-\epsilon^{p/2})^2\|A\|_{S_p}^p = (1-2\epsilon^{p/2} + \epsilon^p)\|A\|_{S_p}^p.
\]
By Lemma \ref{fact:SchattenIneqForSum},
\[
\|A^TA\|_{S_{p/2}}^{p/2}= \|X^TX+Y^TY\|_{S_{p/2}}^{p/2}\geq \|X^TX\|_{S_{p/2}}^{p/2} + \|Y^TY\|_{S_{p/2}}^{p/2}.
\]
Hence 
\begin{equation}\label{eq:large_p:X_upper_bound}
    \|X\|_{S_{p}}^p\leq (2\epsilon^{p/2} - \epsilon^p) \|A\|_{S_{p}}^p\leq (2\epsilon^{p/2} - \epsilon^p) \|A^\top A + C^\top C\|_{S_{p/2}}^{p/2} 
\end{equation}
for any real matrix $C$. By the definition of $A$, triangle inequality and equation \ref{eq:large_p:X_upper_bound},
\begin{align*}
\|A^\top A + C^\top C\|_{S_{p/2}} 
&= \|X^\top X + Y^\top Y + C^\top C\|_{S_{p/2}} \\
&\leq 
\|Y^\top Y + C^\top C\|_{S_{p/2}} + \|X^\top X\|_{S_{p/2}} \\
&\leq \|Y^\top Y + C^\top C\|_{S_{p/2}} +  (2\epsilon^{p/2} - \epsilon^p)^{2/p}\|A^\top A +C^\top C\|_{S_{p/2}}.
\end{align*}
Hence,
\[
\|Y^\top Y + C^\top C\|_{S_{p/2}} \geq \big(1-(2\epsilon^{p/2} - \epsilon^p)^{2/p}\big)\|A^\top A +C^\top C\|_{S_{p/2}}.
\]
By Lemma \ref{fact:epsBoundForLargeP}, $\sqrt{1-(2\epsilon^{p/2} - \epsilon^p)^{2/p}}\geq 1-\epsilon$, which concludes the proof of Proposition \ref{lem:p>2}.
\end{proof}

\section{Smoothness of Schatten Norms for $p<2$}

In this section, we prove Proposition \ref{lem:p<2}, that Schatten $p$-norms for $p\leq 2$ in row-order streams are $(\epsilon,\epsilon)$-smooth.
As in the $p\geq 2$ case, recall that \citet{krauthgamer2019almost} showed that Schatten norms are non-negative, non-decreasing and bounded. 
The remaining part of the proof is based on the proof for $\ell_q$ frequency moments for $q\leq 1$ \citep[Lemma 4]{braverman2007smooth}.

To prove Proposition \ref{lem:p<2}, we will need an auxiliary lemma.
It is well known that if a function $f:\R\rightarrow \R$ is monotonically decreasing, then every two positive semidefinite (PSD) matrices $A\succeq B$ satisfy $\Tr[f(A)]\leq \Tr[f(B)]$. 
We need an analogous statement for the matrix function $f:X\mapsto (X+C^\top C)^q-X^q$ defined for PSD matrices $X$, where $C^\top C$ is a fixed PSD matrix.
While the monotonicity of the trace function does not directly apply here,
the desired monotonicity still holds, as summarized in the next lemma.

\begin{lemma}\label{lem:traceDecreasing}
Let $C\in\R^{n\times m}$. For all $0<q<1$, if $A\succeq B\succeq 0$ then $\Tr[(A+C^\top C)^q-A^q]\leq \Tr[(B+C^\top C)^q-B^q]$.
\end{lemma}
\begin{proof}
Denote the operator function $f(X)=(X+C^\top C)^q-X^q$ for $X\succeq 0$.
Define $$g(t) = \Tr \big[f\big(B+t(A-B)\big)\big] = \Tr\big[\big(B+t(A-B)+C^\top C\big)^q-\big(B+t(A-B)\big)^q\big].$$
Its derivative for $t\in[0,1]$ is
\begin{align*}
  g'(t) & = q\Tr\big[\big(B+t(A-B)+C^\top C\big)^{q-1}(A-B)-\big(B+t(A-B)\big)^{q-1}(A-B)\big]\\
  & = q\Tr\Big[(A-B)^{1/2}\Big(\big(B+t(A-B)+C^\top C\big)^{q-1}-\big(B+t(A-B)\big)^{q-1}\Big)(A-B)^{1/2}\Big]\leq 0,
\end{align*}
where the last step is since $(B+t(A-B)+C^\top C)^{q-1}\prec (B+t(A-B))^{q-1}$ for $q<1$, so the matrix inside the trace is negative semidefinite.
Hence, $\Tr[f(A)]-\Tr[f(B)]=\int_0^1 g'(t)dt\leq 0$.
\end{proof}

\begin{proof}[Proof of Proposition \ref{lem:p<2}.]
Schatten norms are non-negative, non-decreasing and bounded \citep[Corollary 3.9]{krauthgamer2019almost}. 
Let $p\leq 2$ and let $X,Y,C$ be real matrices as in Definition \ref{def:smoothness}, such that $A=\big[\begin{smallmatrix}X \\ Y\end{smallmatrix}\big]$ satisfies 
\begin{equation}\label{eq:small_p:AY_ineq}
    (1-\epsilon)\|A\|_{S_p}\leq \|Y\|_{S_p}. 
\end{equation}
Note that
\begin{equation}\label{eq:small_p:schatten_to_trace}
\bigg\|\bigg[\begin{matrix}
Y \\ C
\end{matrix}\bigg]\bigg\|_{S_p}^p - \|Y\|_{S_p}^p
 =  \Tr \big[(Y^\top Y +C^\top C)^{p/2}-(Y^\top Y)^{p/2}\big].
\end{equation}

Now, by Lemma \ref{lem:traceDecreasing},
and since $A^\top A = X^\top X+Y^\top Y \succeq Y^\top Y$,  then
\[\Tr \big[(Y^\top Y +C^\top C)^{p/2}-(Y^\top Y)^{p/2}\big] 
\geq \Tr \big[(A^\top A +C^\top C)^{p/2}-(A^\top A)^{p/2}\big].\]
Combining this with equations \ref{eq:small_p:AY_ineq} and \ref{eq:small_p:schatten_to_trace}, we get the desired result,
\[
\bigg\|\bigg[\begin{matrix}
Y \\ C
\end{matrix}\bigg]\bigg\|_{S_p}^p \geq 
(1-\epsilon)^{p}\|A\|_{S_p}^p + (1-\epsilon)^{p}\Tr [(A^\top A +C^\top C)^{p/2}-(A^\top A)^{p/2}]
= (1-\epsilon)^{p}
\bigg\|\bigg[\begin{matrix}
A \\ C
\end{matrix}\bigg]\bigg\|_{S_p}^p.
\]
\end{proof}

\paragraph{Acknowledgements}
This work was partially supported by ONR Award N00014-18-1-2364, the Israel Science Foundation grant \#1086/18, and a Minerva Foundation grant.

\bibliographystyle{elsarticle-num-names} 
\bibliography{references.bib}

\begin{thebibliography}{16}
\expandafter\ifx\csname natexlab\endcsname\relax\def\natexlab#1{#1}\fi
\providecommand{\url}[1]{\texttt{#1}}
\providecommand{\href}[2]{#2}
\providecommand{\path}[1]{#1}
\providecommand{\DOIprefix}{doi:}
\providecommand{\ArXivprefix}{arXiv:}
\providecommand{\URLprefix}{URL: }
\providecommand{\Pubmedprefix}{pmid:}
\providecommand{\doi}[1]{\href{http://dx.doi.org/#1}{\path{#1}}}
\providecommand{\Pubmed}[1]{\href{pmid:#1}{\path{#1}}}
\providecommand{\bibinfo}[2]{#2}
\ifx\xfnm\relax \def\xfnm[#1]{\unskip,\space#1}\fi
%Type = Inproceedings
\bibitem[{Liberty(2013)}]{liberty2013simple}
\bibinfo{author}{E.~Liberty},
\newblock \bibinfo{title}{Simple and deterministic matrix sketching},
\newblock in: \bibinfo{booktitle}{Proceedings of the 19th ACM SIGKDD
  International Conference on Knowledge Discovery and Data Mining},
  \bibinfo{year}{2013}, pp. \bibinfo{pages}{581--588}.
  \DOIprefix\doi{10.1145/2487575.2487623}.
%Type = Article
\bibitem[{Datar et~al.(2002)Datar, Gionis, Indyk, and
  Motwani}]{datar2002maintaining}
\bibinfo{author}{M.~Datar}, \bibinfo{author}{A.~Gionis},
  \bibinfo{author}{P.~Indyk}, \bibinfo{author}{R.~Motwani},
\newblock \bibinfo{title}{Maintaining stream statistics over sliding windows},
\newblock \bibinfo{journal}{SIAM Journal on Computing} \bibinfo{volume}{31}
  (\bibinfo{year}{2002}) \bibinfo{pages}{1794--1813}.
  \DOIprefix\doi{10.1137/S0097539701398363}.
%Type = Inproceedings
\bibitem[{Wei et~al.(2016)Wei, Liu, Li, Shang, Du, and Wen}]{wei2016matrix}
\bibinfo{author}{Z.~Wei}, \bibinfo{author}{X.~Liu}, \bibinfo{author}{F.~Li},
  \bibinfo{author}{S.~Shang}, \bibinfo{author}{X.~Du},
  \bibinfo{author}{J.~Wen},
\newblock \bibinfo{title}{Matrix sketching over sliding windows},
\newblock in: \bibinfo{booktitle}{Proceedings of the 2016 International
  Conference on Management of Data}, \bibinfo{publisher}{{ACM}},
  \bibinfo{year}{2016}, pp. \bibinfo{pages}{1465--1480}.
  \DOIprefix\doi{10.1145/2882903.2915228}.
%Type = Inproceedings
\bibitem[{Braverman et~al.(2020)Braverman, Drineas, Musco, Musco, Upadhyay,
  Woodruff, and Zhou}]{braverman2020near}
\bibinfo{author}{V.~Braverman}, \bibinfo{author}{P.~Drineas},
  \bibinfo{author}{C.~Musco}, \bibinfo{author}{C.~Musco},
  \bibinfo{author}{J.~Upadhyay}, \bibinfo{author}{D.~P. Woodruff},
  \bibinfo{author}{S.~Zhou},
\newblock \bibinfo{title}{Near optimal linear algebra in the online and sliding
  window models},
\newblock in: \bibinfo{booktitle}{61st {IEEE} Annual Symposium on Foundations
  of Computer Science, {FOCS}}, \bibinfo{publisher}{{IEEE}},
  \bibinfo{year}{2020}, pp. \bibinfo{pages}{517--528}.
%Type = Article
\bibitem[{Khetan and Oh(2019)}]{khetan2019spectrum}
\bibinfo{author}{A.~Khetan}, \bibinfo{author}{S.~Oh},
\newblock \bibinfo{title}{Spectrum estimation from a few entries},
\newblock \bibinfo{journal}{The Journal of Machine Learning Research}
  \bibinfo{volume}{20} (\bibinfo{year}{2019}) \bibinfo{pages}{21:1--21:55}.
  \URLprefix \url{http://jmlr.org/papers/v20/18-027.html}.
%Type = Article
\bibitem[{Kong and Valiant(2017)}]{kong2017spectrum}
\bibinfo{author}{W.~Kong}, \bibinfo{author}{G.~Valiant},
\newblock \bibinfo{title}{{Spectrum estimation from samples}},
\newblock \bibinfo{journal}{The Annals of Statistics} \bibinfo{volume}{45}
  (\bibinfo{year}{2017}) \bibinfo{pages}{2218 -- 2247}.
  \DOIprefix\doi{10.1214/16-AOS1525}.
%Type = Inproceedings
\bibitem[{Nie et~al.(2012)Nie, Huang, and Ding}]{nie2012low}
\bibinfo{author}{F.~Nie}, \bibinfo{author}{H.~Huang},
  \bibinfo{author}{C.~Ding},
\newblock \bibinfo{title}{Low-rank matrix recovery via efficient {S}chatten
  $p$-norm minimization},
\newblock in: \bibinfo{booktitle}{Proceedings of the Twenty-Sixth AAAI
  Conference on Artificial Intelligence}, \bibinfo{publisher}{AAAI Press},
  \bibinfo{year}{2012}, p. \bibinfo{pages}{655–661}. \URLprefix
  \url{http://www.aaai.org/ocs/index.php/AAAI/AAAI12/paper/view/5165}.
%Type = Inproceedings
\bibitem[{Li et~al.(2014)Li, Nguyen, and Woodruff}]{li2014sketching}
\bibinfo{author}{Y.~Li}, \bibinfo{author}{H.~L. Nguyen}, \bibinfo{author}{D.~P.
  Woodruff},
\newblock \bibinfo{title}{On sketching matrix norms and the top singular
  vector},
\newblock in: \bibinfo{booktitle}{Proceedings of the Twenty-Fifth Annual
  {ACM-SIAM} Symposium on Discrete Algorithms, {SODA}},
  \bibinfo{publisher}{{SIAM}}, \bibinfo{year}{2014}, pp.
  \bibinfo{pages}{1562--1581}. \DOIprefix\doi{10.1137/1.9781611973402.114}.
%Type = Inproceedings
\bibitem[{Li and Woodruff(2016{\natexlab{a}})}]{li2016approximating}
\bibinfo{author}{Y.~Li}, \bibinfo{author}{D.~P. Woodruff},
\newblock \bibinfo{title}{On approximating functions of the singular values in
  a stream},
\newblock in: \bibinfo{booktitle}{Proceedings of the 48th Annual {ACM} {SIGACT}
  Symposium on Theory of Computing, {STOC} 2016}, \bibinfo{publisher}{{ACM}},
  \bibinfo{year}{2016}{\natexlab{a}}, pp. \bibinfo{pages}{726--739}.
  \DOIprefix\doi{10.1145/2897518.2897581}.
%Type = Inproceedings
\bibitem[{Li and Woodruff(2016{\natexlab{b}})}]{li2016tight}
\bibinfo{author}{Y.~Li}, \bibinfo{author}{D.~P. Woodruff},
\newblock \bibinfo{title}{Tight bounds for sketching the operator norm,
  {S}chatten norms, and subspace embeddings},
\newblock in: \bibinfo{booktitle}{Approximation, Randomization, and
  Combinatorial Optimization. Algorithms and Techniques, {APPROX/RANDOM}},
  \bibinfo{year}{2016}{\natexlab{b}}, pp. \bibinfo{pages}{39:1--39:11}.
  \DOIprefix\doi{10.4230/LIPIcs.APPROX-RANDOM.2016.39}.
%Type = Inproceedings
\bibitem[{Braverman et~al.(2018)Braverman, Chestnut, Krauthgamer, Li, Woodruff,
  and Yang}]{braverman2018matrix}
\bibinfo{author}{V.~Braverman}, \bibinfo{author}{S.~Chestnut},
  \bibinfo{author}{R.~Krauthgamer}, \bibinfo{author}{Y.~Li},
  \bibinfo{author}{D.~Woodruff}, \bibinfo{author}{L.~Yang},
\newblock \bibinfo{title}{Matrix norms in data streams: Faster, multi-pass and
  row-order},
\newblock in: \bibinfo{booktitle}{International Conference on Machine
  Learning}, \bibinfo{organization}{PMLR}, \bibinfo{year}{2018}, pp.
  \bibinfo{pages}{649--658}. \URLprefix
  \url{http://proceedings.mlr.press/v80/braverman18a.html}.
%Type = Inproceedings
\bibitem[{Braverman et~al.(2020)Braverman, Krauthgamer, Krishnan, and
  Sinoff}]{braverman2020schatten}
\bibinfo{author}{V.~Braverman}, \bibinfo{author}{R.~Krauthgamer},
  \bibinfo{author}{A.~Krishnan}, \bibinfo{author}{R.~Sinoff},
\newblock \bibinfo{title}{Schatten norms in matrix streams: Hello sparsity,
  goodbye dimension},
\newblock in: \bibinfo{booktitle}{Proceedings of the 37th International
  Conference on Machine Learning, {ICML}}, \bibinfo{organization}{PMLR},
  \bibinfo{year}{2020}, pp. \bibinfo{pages}{1100--1110}. \URLprefix
  \url{http://proceedings.mlr.press/v119/braverman20b.html}.
%Type = Article
\bibitem[{Krauthgamer and Reitblat(2019)}]{krauthgamer2019almost}
\bibinfo{author}{R.~Krauthgamer}, \bibinfo{author}{D.~Reitblat},
\newblock \bibinfo{title}{Almost-smooth histograms and sliding-window graph
  algorithms},
\newblock \bibinfo{journal}{arXiv preprint arXiv:1904.07957}
  (\bibinfo{year}{2019}).
%Type = Inproceedings
\bibitem[{Braverman and Ostrovsky(2007)}]{braverman2007smooth}
\bibinfo{author}{V.~Braverman}, \bibinfo{author}{R.~Ostrovsky},
\newblock \bibinfo{title}{Smooth histograms for sliding windows},
\newblock in: \bibinfo{booktitle}{48th Annual {IEEE} Symposium on Foundations
  of Computer Science {(FOCS} 2007)}, \bibinfo{publisher}{{IEEE} Computer
  Society}, \bibinfo{year}{2007}, pp. \bibinfo{pages}{283--293}.
  \DOIprefix\doi{10.1109/FOCS.2007.55}.
%Type = Article
\bibitem[{Woodruff and Zhou(2020)}]{DBLP:journals/arxiv/WoodruffZ20}
\bibinfo{author}{D.~P. Woodruff}, \bibinfo{author}{S.~Zhou},
\newblock \bibinfo{title}{Tight bounds for adversarially robust streams and
  sliding windows via difference estimators},
\newblock \bibinfo{journal}{arXiv preprint arXiv:2011.07471}
  (\bibinfo{year}{2020}). \bibinfo{note}{To appear in FOCS 2021}.
%Type = Book
\bibitem[{Bhatia(1997)}]{bhatia1997matrix}
\bibinfo{author}{R.~Bhatia}, \bibinfo{title}{Matrix analysis}, volume
  \bibinfo{volume}{169} of \textit{\bibinfo{series}{Graduate Texts in
  Mathematics}}, \bibinfo{publisher}{Springer-Verlag, New York},
  \bibinfo{year}{1997}. \DOIprefix\doi{10.1007/978-1-4612-0653-8}.

\end{thebibliography}

\end{document}